\documentclass[superscriptaddress, nofootinbib, preprintnumbers, twocolumn]{revtex4}

\usepackage{amsmath}	
\usepackage{amsthm}		
\usepackage{amssymb}	
\usepackage{eufrak}
\usepackage{datetime}
\usepackage{graphicx}
\usepackage{color}
\usepackage{verbatim}
\usepackage{bm}

\usepackage[colorlinks=true, citecolor=midblue, linkcolor=midblue, urlcolor=midblue]{hyperref}

\usepackage{setspace}

\definecolor{grey}{rgb}{0.4,0.4,0.4}
\definecolor{dullmagenta}{rgb}{0.4,0,0.4}
\definecolor{darkblue}{rgb}{0,0,0.4}
\definecolor{midblue}{rgb}{0,0,0.5}
\definecolor{midred}{rgb}{0.5,0,0}
\definecolor{orange}{rgb}{1,0.5,0}
\definecolor{lightbrown}{rgb}{0.75,0.5,0.25}
\definecolor{tan}{cmyk}{0.14,0.42,0.56,0}
\definecolor{djunglegreen}{cmyk}{0.99,0,0.52,0}
\definecolor{lightgreen}{rgb}{0,1,0}
\definecolor{olivegreen}{cmyk}{0.64,0,0.95,0.40}
\definecolor{midgreen}{rgb}{0.0,0.675,0.0}
\definecolor{darkgreen}{rgb}{0,0.5,0}

\newcommand{\vs}{\vspace}

\newcommand{\Irm}{\ensuremath{\mathrm{I}}}

\newcommand{\Lrm}{\ensuremath{\mathrm{L}}}

\newcommand{\Prm}{\ensuremath{\mathrm{P}}}

\newcommand{\Rrm}{\ensuremath{\mathrm{R}}}

\newcommand{\Yrm}{\ensuremath{\mathrm{Y}}}

\newcommand{\crm}{\ensuremath{\mathrm{c}}}
\newcommand{\drm}{\ensuremath{\mathrm{d}}}
\newcommand{\erm}{\ensuremath{\mathrm{e}}}

\newcommand{\hrm}{\ensuremath{\mathrm{h}}}

\newcommand{\srm}{\ensuremath{\mathrm{s}}}
\newcommand{\trm}{\ensuremath{\mathrm{t}}}

\newcommand{\RH}{{\rm RH}}
\newcommand{\QCD}{{\rm QCD}}
\newcommand{\PQ}{{\rm PQ}}
\newcommand{\eff}{{\rm eff}}
\newcommand{\GH}{{\rm GH}}
\newcommand{\Inf}{{\rm Inf}}

\newcommand{\GeV}{\ensuremath{\, \mathrm{GeV}}}

\smallskipamount

\settimeformat{ampmtime}

\usepackage{float}

\begin{document}

\title{Removing the Cosmological Bound on the Axion Scale in the KSVZ and DFSZ Models}

\author{Emmanouil Koutsangelas}
\email{emi@mpp.mpg.de}
\affiliation{
	Arnold Sommerfeld Center,
	Ludwig-Maximilians-Universit{\"a}t,
	Theresienstra{\ss}e 37,
	80333 M{\"u}nchen,
	Germany,}
\affiliation{
	Max-Planck-Institut f{\"u}r Physik,
	F{\"o}hringer Ring 6,
	80805 M{\"u}nchen,
	Germany}

\date{\formatdate{\day}{\month}{\year}
}

\begin{abstract}
	It has been known for some time that the cosmological bound on the invisible axion scale can be avoided by an early phase of strong QCD. While most approaches rely on theories where the strong coupling constant is determined through the expectation value of some scalar field, we show that this mechanism can also be implemented into the benchmark KSVZ and DFSZ models where the early phase of strong QCD emerges by the modification of the running coupling during inflation. For both models the physics that are responsible for making QCD strong do not displace the axions minimum by too much, so that the efficiency of the relaxation is controlled by parameters of the theory and the number of inflationary $e$-folds. In particular, we consider the case of very efficient relaxation where the axion abundance is dominated by inflationary quantum and postinflationary thermal fluctuations. Within this situation we identify the parameter space compatible with all cosmological constrains and  derive conditions on the reheating temperature and the QCD scale during inflation that result in the axion making up all the dark matter. Because of duality below the Peccei-Quinn scale and a minor influence of the KSVZ and DFSZ fields on the running, our findings also apply to a minimal two-form implementation of the axion.
\end{abstract}

\maketitle

\tableofcontents

\section{Introduction}
\label{sec:Introduction}
\vs{-5mm}

The axion is one of the best motivated particle candidates beyond the Standard Model. It does not only solve the strong $CP$ problem but due to its Goldstone nature it also provides an excellent dark matter candidate. Because of that, the axion is of high interest not only to particle physics but also to other branches of physics such as cosmology and astrophysics. Therefore, it is unsurprising that today the axion experimental program experiences a time of prosperity where regions of the axion parameter space are explored that were considered science fiction until a decade ago.

Clearly, this makes an understanding of the parameter space of the axion indispensable. Let us put the focus on the region with large values of the axion decay constant, $f_a \gtrsim 10^{12} \GeV$, which is usually excluded from the axion window because of overproduction from the misalignment mechanism \cite{MisalignmentPRESKILL, MisalignmentDINE, MisalignmentAbbott}. If this bound was taken as cast in stone, a nonmeasurement of the axion in the classical window $10^{9} \GeV \lesssim f_a \lesssim 10^{12} \GeV$ would result in its exclusion. Alternatively, a measurement at higher values of $f_a$ would be interpreted as an axionlike particle instead of the QCD axion. But if there was a way to avoid this cosmological bound, the interpretation would be vastly different. This shows the importance of understanding imposed bounds and, particularly, the possibility to avoid them. 

The goal of this work is to do exactly that, namely, to extend on Dvali's scenario \cite{Dvali:1995ce} and provide evidence on the nonrobustness of the mentioned cosmological bound. In fact, the bound is based on the crucial assumption of the initial misalignment angle $\theta_1$ being of $\mathcal{O}(1)$ or, in other words, being not fine-tuned. While fine-tuning is, per se, not a problem, it is regarded as unattractive in the absence of an explanation of its origin. Hence, if there was a mechanism that results in very small values of $\theta_1$, the region with $f_a \gtrsim 10^{12} \GeV$ would be rendered natural. 

This would be of particular interest for string axionlike particles \cite{Arvanitaki:2009fg} or even the pure gauge two-form implementation of the axion \cite{dvali2005threeform, PhysRevD.105.085020} since both favor large values of $f_a$. As explained in \cite{Dvali:2022fdv}, the latter is motivated by gravity, which not only demands the axion mechanism exact to all orders in operator expansion but also connects $f_a$ to the gravitational cutoff. This also refines the original constraint of \cite{Dvali:2018dce} about the necessity of the axion in gravity. 

The general idea of the mechanism proposed in \cite{Dvali:1995ce} is based on a temporal phase of strong QCD at some time well before the misalignment mechanism normally takes place. In that case the axion develops its potential in the early Universe and, provided that $m_a \gtrsim H$ at that time or the phase of strong QCD lasts long enough, relaxes to its minimum. At the end of this early phase of strong QCD the potential then vanishes, leaving the axion frozen out at its particular value at that time. When the ``ordinary" QCD phase transition activates the axion potential again, the axion is already located close to the minimum. The resulting amplitude of the oscillations in the misalignment mechanism is then strongly suppressed, which is equivalent to the misalignment angle being naturally small. 

The point made in \cite{Dvali:1995ce} is that this situation is especially characteristic to inflation, which is fundamentally based on the idea that fields are displaced from their today's values in the early Universe. In this way the inflaton is expected to affect the values of the couplings and vacuum expectation values (VEVs) in the Standard Model. Particularly, as argued there, the QCD coupling could become strong resulting into the generation of the axion potential in the early Universe.

In this work, we study how this mechanism can be realized in the two benchmark invisible axion models, namely, the KSVZ \cite{KSVZ1, KSVZ2} and DFSZ \cite{DFSZ1, DFSZ2} models. Our findings show that a larger VEV of the Higgs doublet during inflation and the associated larger quark masses modify the value of the QCD scale sufficiently to achieve a strong phase of QCD for a large range of values of the inflationary Hubble parameter. Such a shift in the Higgs VEV can be generated via a simple Higgs portal between our Higgs doublet and the inflaton or higher dimensional operators of those fields. In the KSVZ and the DFSZ models the physics that make QCD strong in the early Universe do not change the location of the axions minimum. Therefore, the axions minimum during inflation essentially coincides with its late time minimum and the mechanism is expected to work. The amount of suppression of the misalignment angle within these circumstances is then controlled by the number of inflationary $e$-folds, the axion mass, and the Hubble parameter during inflation.

An interesting point to make is that our analysis is automatically valid for the mentioned two-form implementation of the axion \cite{Dvali:2022fdv}. The relevant point of this implementation for us is that below the PQ scale the two-form axion is exactly dual to the pseudoscalar axion. Thus, from this point of view our analysis does automatically apply to the two-form implementation. The two-form implementation, however, does not make any statement about the additional field content [such as additional KSVZ, DFSZ, or grand unified theory (GUT) fields] that alters the running of the QCD gauge coupling. Of course, we cannot exclude the existence of those additional fields from nonaxion related reasons. But as we demonstrate, the modification of the running by the additional fields in the benchmark models is negligible so that our analysis is valid for a minimal two-form implementation.

In principle, the initial misalignment angle can take values so small that the axion energy density is controlled by inflationary quantum fluctuations and/or postinflationary thermal fluctuations. In this extreme case, constrains coming from isocurvature fluctuations and inflationary gravitational waves generically result in a relatively narrow region of the parameter space becoming viable. The two models at hand are not compatible with this extreme case; hence, it provides a bound on the number of inflationary $e$-folds, even though the bound is not very restrictive. It should be stressed, however, that the described mechanism is not constrained to the benchmark KSVZ and DFSZ models. It can also be implemented into general KSVZ- or DFSZ-type models (see for instance \cite{DiLuzio:2017pfr, Plakkot:2021xyx} and \cite{Diehl:2023uui} for catalogs of preferred KSVZ- and DFSZ-type models, respectively), which on the other hand could be embedded into GUT models. The field content is different in these models so that the running of the QCD coupling constant and thus the value of the QCD scale during inflation are altered, possibly resulting in the extreme case.

Our work is supplementing to other studies of Dvali's early relaxation mechanism that have been performed in the literature. For instance, see \cite{Choi:1996fs} and \cite{Co:2018phi} for an implementation in the SUSY context. In \cite{Matsui:2020wfx} the implications from Higgs inflation were studied. For the related idea of low-scale inflation which focuses on building a concrete QCD-scale inflation model with successful reheating, see \cite{Guth:2018tdu}. A list of alternative ideas to enlarge the axion window can be found in \cite{DiLuzioAxionLandscape}. Let us also mention that a complementary study of the same mechanism was given in \cite{AAdI}.

This paper is structured as follows. To begin with, in Sec. \ref{sec:Cosmological_Bounds} we briefly review the cosmological bounds on the axion where the focus lies on the cosmological production via the misalignment mechanism. In particular, we highlight the role of the initial misalignment angle $\theta_1$ together with the implications coming from thermal and inflationary fluctuations important for our final result. In Sec. \ref{sec:Early_Relaxation_using_Inflation} we study how an early relaxation can be realized. First we define the models under consideration. Next, we show how in these models the QCD scale is shifted during inflation and how an early phase of strong QCD can emerge. We then calculate the initial misalignment angle in the KSVZ and DFSZ models using the inflationary induced shift on the axions minimum. Lastly, we calculate the energy density and specify the region of the parameter space that becomes viable in our model. In the final section, we then summarize and give an outlook.

\section{Cosmological Bounds on the Axion}
\label{sec:Cosmological_Bounds}
\vs{-5mm}

\subsection{Production via misalignment}
\label{subsec:Overpopulating_the_Universe}
\vs{-5mm}

The standard misalignment mechanism for the axion works as follows. At the temperature $T_\PQ \sim v_a$ the U(1)$_\PQ$ symmetry is spontaneously broken and the axion $a(x)$ emerges as the corresponding Goldstone boson \cite{PQMechanism, WeinbergAxion, WilczekAxion}. Using the standard notation, we canonically normalize the Goldstone field as $\theta(x) \equiv 2 N a(x)/v_a \equiv a(x)/f_a$, where $f_a$ is the axion decay constant and $N$ is the QCD anomaly coefficient.

Because of the Goldstone nature of the axion, its potential is initially flat. Eventually, the temperature becomes of order of the QCD scale $\Lambda_\QCD\sim0.2 \GeV$ and then nonperturbative QCD effects generate an effective potential for the axion, which breaks the initial U(1) symmetry down to the discrete subgroup $\mathbb{Z}_{2 N}$. To the leading order in the semiclassical approximation these effects can be accounted for by instantons that are coupled to a thermal bath. Using the dilute instanton gas approximation \cite{PhysRevD.17.2717}, the temperature-dependent potential takes the form \cite{QCDINstantonsFiniteTemp}
	\begin{equation}
		V(\theta)	
			=
				m_a^2(T) f_a^2
				\big(1 - \cos(\theta)\big)
			\; ,
	\label{Eq:PotDIGA}
	\end{equation}
where for the axion mass we use
	\begin{equation}
		m_a(T)
			\equiv
				\frac{(\Lambda_\QCD^3 m_u)^{\frac{1}{2}}}{f_a}
				\begin{cases}
					\gamma
					\left(
						\frac{\Lambda_\QCD}{T}
					\right)^4 
					&:T > \Lambda_\QCD
					\; ,
						\\
					1 
					&:T \lesssim \Lambda_\QCD
					\; .
				\end{cases}
	\label{Eq:AxionMassInstanonThermal}
	\end{equation}
Here, $m_u$ denotes the mass of the up-quark and $\gamma$ encodes QCD and active quark physics, which for the Standard Model roughly are of order $10^{-2}$ \cite{PhysRevD.17.2717}. 

The cosmological behavior of the axion field is dictated by its equation of motion in the Friedmann–Lemaître–Robertson–Walker background, 
	\begin{equation}
		\ddot{\theta} 
		+ 3H(t) \dot{\theta} 
		- \frac{1}{R^2(t)} \Delta \theta 
		+ \frac{V'(\theta)}{f_a^2} 
			= 
				0
		\; ,
	\label{Eq:EomTheta}
	\end{equation}
where $R(t)$ denotes the scale factor and $H(t)$ the Hubble parameter. Let us use the potential \eqref{Eq:PotDIGA} and make the following two simplifications. First, since we are interested in small values of $\theta$ we can neglect higher orders in the expansion of the potential. Second, let us only consider the zero mode of $\theta$. With these simplifications, the equation of motion reduces to that of a damped harmonic oscillator,
	\begin{equation}
		\ddot{\theta} 
		+ 3H(t) \dot{\theta} 
		+ m_a^2\big( T(t) \big) \theta
			= 
				0
			\; .
	\label{Eq:KGTheta}
	\end{equation}
At $T >> \Lambda_\QCD$ the mass is essentially zero so that the Hubble friction dominates and the axion is essentially frozen out. Eventually, $m_a \big( T(t) \big) \sim H(t)$, so the axion can start performing coherent oscillations around the vacuum. From that moment on, the axion's equation of state no longer corresponds to that of dark energy but to that of nonrelativistic matter, making the axion contribute into the dark matter. 

Let us use the Universe's temperature instead of cosmic time and define this moment by $m_a(T_1) = 3 H(T_1)$. In order to calculate $T_1$, we use that this equality occurs during radiation domination when the Hubble parameter is given by
	\begin{equation}
		H (T)
			=
				\sqrt{\frac{\rho}{3 M_\Prm^2}}
			\sim
				\frac{ T^2 }
				{M_\Prm}
			\; .
	\end{equation}
Combining this with the axion mass defined by the first row of (\ref{Eq:AxionMassInstanonThermal}), this results in the oscillations commencing at
	\begin{equation}
		T_1
			\sim
				1 \GeV \left(
				\frac{10^{12} \GeV}{f_a}
				\right)^\frac{1}{6}
			\; .
		\label{Eq:T1}
	\end{equation}
Note that for values of $f_a \gtrsim 2 \times 10^{17} \GeV$ the oscillations begin when the axion has already reached its low-temperature mass, which results in a different expression for $T_1$. Furthermore, in the region where $2 \times 10^{15} \GeV \lesssim f_a \lesssim 2 \times 10^{17} \GeV$ there is a slight caveat due to the breakdown of the dilute instanton gas approximation. Let us nevertheless use (\ref{Eq:T1}) for all $f_a < M_\Prm$ for the purposes of illustration.

The contribution to today's dark matter fraction can be estimated as follows. The initial energy density of these oscillations at $T_1$ is
	\begin{equation}
		\rho_1
			= 
				\frac{1}{2} f_a^2 
				m_a^2(T_1) \theta_1^2
			\; ,
	\end{equation}
where $\theta_1$ is the initial misalignment angle. Since $\theta_1$ is the value of the canonically normalized axion when the oscillations begin, it is nothing other than the amplitude at that particular moment.

Since the axions are decoupled, the number of zero modes per co-moving volume is conserved as long as the changes in the mass are in the adiabatic regime. Assuming this to be the case, today's energy density is given by
	\begin{equation}
		\rho_0 
			= 
				\rho_1 
				\frac{m_a(T_0)}{m_a(T_1)}
				\left(
					\frac{R(T_1)}{R(T_0)}
				\right)^3
			\; ,
	\end{equation}
and the associated cold dark matter fraction is
	\begin{equation}
		\frac{\Omega_a }{\Omega_{\rm CDM}}
			\sim
				\left(
					\frac{f_a}{10^{12} \GeV}
				\right)^{7/6} 
				\theta_1^2
				\; .
	\label{Eq:DMfraction}
	 \end{equation}
This simple estimate shows that there is an upper bound on $f_a$ depending on the value of $\theta_1$. For non-fine-tuned values of the initial misalignment angle, i.e., $\theta_1 \sim \mathcal{O}(1)$, we roughly have $f_a \lesssim 10^{12}$ GeV.

It is important to distinguish between two scenarios that are determined by the time when the PQ symmetry is broken. Let us introduce the following scales in order to quantify these two scenarios:
	\begin{itemize}
		\item The Gibbons-Hawking temperature during inflation, $T_\GH = H_\Irm/2\pi$, where $H _\Irm$ is the inflationary Hubble parameter.
		\item The maximum thermalization temperature \cite{Hertzberg:2008wr}
			\begin{equation}
				T_{\rm Max}
					= 
						\epsilon_\eff 
						E_\Irm 
					\sim 
						\epsilon_\eff
						(M_\Prm H_\Irm)^{\frac{1}{2}}
					\; ,
			\label{Eq:TMax}
			\end{equation}
		where $\epsilon_\eff$ is used as a dimensionless efficiency parameter with $0 \lesssim \epsilon_\eff \lesssim 1 $. This should not be confused with the reheating temperature which can be somewhat lower \cite{Kolb:2003ke}.
	\end{itemize}

The preinflationary scenario takes place when the PQ symmetry is spontaneously broken during or before inflation and never restored afterwards, i.e., $f_a~>~\max~\{T_{\rm GH}, T_{\rm RH} \}$. Since the axion has no potential in the beginning, in each Hubble patch one initial value is chosen randomly from a uniform distribution on the range $[-\pi, \pi]$. Inflation then enlarges each patch so that after inflation has ended the Universe starts with one homogeneous value $\theta_1$. This makes $\theta_1$ a free parameter in this scenario.

\begin{figure}[t]
	\includegraphics[scale=0.67,angle=0]{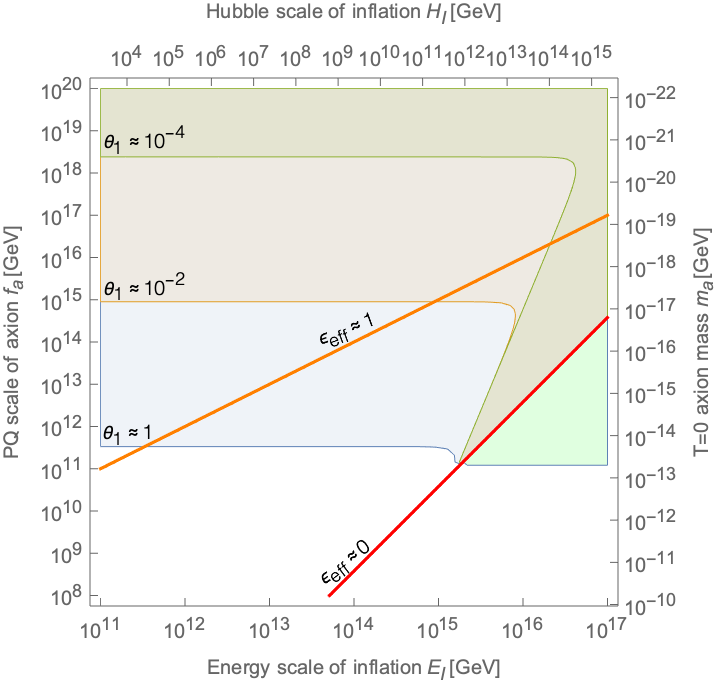}
	\caption{
		Cosmological bound on the axion scale $f_a$ depending on the inflationary Hubble scale $H_\Irm$. The thick red line representing $f_a = \max \{T_{\rm GH}, T_{\rm RH} \}$ in the case of inefficient reheating separates the preinflationary and the postinflationary scenarios. The case of very efficient reheating with $\epsilon_\eff \approx 0$ is also depicted to show the influence of the thermalization efficiency on the bound. We neglect the contribution from the decay of topological defects in the postinflationary scenario, since it plays no role for the purpose of this paper.}
	\label{Fig:CB}
\end{figure}

In contrast, the postinflationary scenario takes place when the PQ symmetry is broken or restored after inflation, i.e., $f_a < \max \{T_{\rm GH}, T_{\rm RH} \}$. In this scenario the patches with different initial values come into causal contact so that there appear all possible values in today's Universe. Thus, the initial misalignment angle in today's universe can be defined as the average over the uniform distribution, i.e., $\theta_1 \equiv \sqrt{\langle \theta^2_1 \rangle} = \pi/\sqrt{3}$. 

It should be mentioned that the two scenarios are different in regard to topological defects such as domain walls and cosmic string. In the preinflationary scenario these defects are inflated away so that they play no role in the further evolution. In the postinflationary scenario, however, these defects are present and thus their decay gives an additional contribution to the energy density of the axion. Since in this paper we will focus on the preinflationary scenario and since there is a large uncertainty in the final  contribution to the axion energy density, we will simply neglect it.

In Fig.~\ref{Fig:CB} we show the cosmological bound for both scenarios. We depict the separation between the two scenarios by a red line representing $f_a = \max \{T_{\rm GH}, T_{\rm Max} \}$, where for concreteness we chose the case of very inefficient reheating ($\epsilon_\eff \approx 0$). Below the red line the postinflationary scenario takes place, while above the red line the preinflationary scenario takes place. In this figure we also depict the line that separates the two scenarios in the other extreme case, namely, very efficient reheating ($\epsilon_\eff \approx 1$), in order to show the dependence on the parameter $\epsilon_\eff$. Moreover, in the preinflationary scenario we included the influence of inflationary fluctuations (see Sec.~\ref{subsec:Isocurvature_Bound}), which are responsible for the deviation from the horizontal lines. 

\subsection{Inflationary quantum fluctuations}
\label{subsec:Isocurvature_Bound}
\vs{-5mm}

In the preinflationary scenario the axion is present during inflation and massless. Hence, it is subject to quantum fluctuations with standard deviation of
	\begin{equation}
		\sigma_{\theta}
			\sim
				\frac{H_\Irm}{2 \pi f_a}
			\; .
	\label{Eq:QuantumFluct}
	\end{equation}
These fluctuations try to move the axion away from its value during inflation, which effectively can be accounted for in (\ref{Eq:DMfraction}) by replacing $\theta_1^2 \rightarrow \theta_1^2 + \sigma_{\theta}^2$.  

Moreover, if the axion is just a spectator field during inflation its fluctuations will not be of adiabatic but of isocurvature type. Since these lead to a unique imprint in the temperature and polarization fluctuations of the CMB, they give rise to a constraint on the axions parameter space.

In the regime of small $\theta_1$ in which anharmonic corrections can be ignored, the amplitude of the axions isocurvature perturbations is given by \cite{DiLuzioAxionLandscape}
	\begin{equation}
		\Delta_{a}^2(k)
			=
				\left(
					\frac{\Omega_{a}}{\Omega_{\rm CDM}}	
				\right)^2
				\left(
					\frac{H_\Irm}{\pi \theta_1 f_a}
				\right)^2
			\; .	
	\end{equation}
The latest experimental bound on uncorrelated isocurvature perturbations by Planck is \cite{Planck2018}
	\begin{equation}
				\frac{ \Delta^2_{a}(k_0))}
				{ \Delta^2_\mathcal{R}(k_0) + \Delta^2_{a}(k_0) }
			<
				 0.038 \quad {\rm at \ 95\% \ CL}
			\; ,
	\end{equation}
where $k_0 = 0.050 {\rm Mpc}^{-1}$. This translates to a constrain on $H_\Irm$,
	\begin{equation}
		H_\Irm
			\lesssim
				10^7 \GeV
				\left(
					\frac{\Omega_{a}}{\Omega_{\rm CDM}}
				\right)^{-1}
				\left(
					\frac{f_a}{10^{12} \GeV}
					\frac{\theta_1}{1}
				\right)
			\; .
	\label{Eq:HIBoundfromISO}
	\end{equation}
	
Finally, let us mention the most robust and model-independent constraint on $H_\Irm$ itself, coming from tensor fluctuations that develop into primordial gravitational waves. Their amplitude $A_\trm = 2 H_\Irm^2/ (\pi^2 M_\Prm^2)$ \cite{Marsh:2015xka} only depends on the expansion rate during inflation and thus represents a vertical bound of the parameter space to the right.
	
It is convenient to express the bound in terms of the tensor-to-scalar ratio $r \equiv A_\trm/A_\srm$, which is just the tensor amplitude normalized with respect to the measured scalar amplitude $A_\srm$. The bound $r < 0.063$ and the value $A_\srm = (2.196 \pm 0.060) \times 10^{-9}$ at $k = 0.05 \, {\rm Mpc}^{-1}$, both given by the Planck-BICEP2 joint analysis \cite{Planck2018}, then translate into
	\begin{equation}
		H_\Irm 
			< 
				6.1 \times 10^{13} \GeV
			\; .
	\label{Eq:GravWaveBound}
	\end{equation}

\begin{figure}[t]
	\includegraphics[scale=0.67,angle=0]{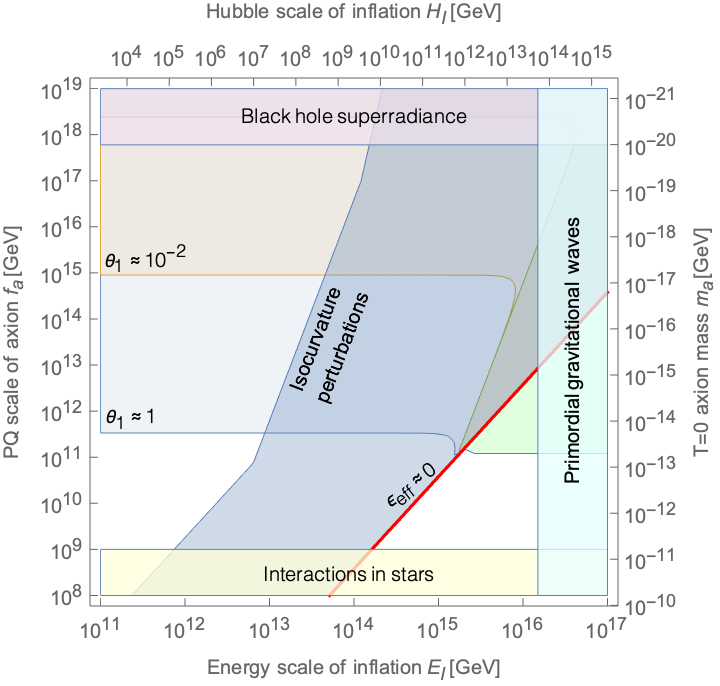}
	\caption{
		Axion window with the bounds coming from isocurvature perturbations (darker blue), black hole superradiance (purple), interactions in stars (yellow), primordial gravitational waves (cyan), and overproduction (light blue or brown, see Fig. \ref{Fig:B} for an isolation of this constraint). The bounds not discussed here are taken from \cite{DiLuzioAxionLandscape}.}
	\label{Fig:B}
\end{figure}

In Fig. \ref{Fig:B} we added the constrain from isocurvature perturbations with the axion assumed to make up all the dark matter. Note that with this requirement, (\ref{Eq:DMfraction}) can be used to eliminate $\theta_1$ in (\ref{Eq:HIBoundfromISO}). Therefore, the dependence on $\theta_1$ is implicit in Fig. \ref{Fig:B}, i.e., higher values of $f_a$ require smaller values of $\theta_1$. We also included the bound on the inflationary Hubble parameter coming from gravitational waves and, for completeness, the other known bounds on the axion (see \cite{DiLuzioAxionLandscape} for details).

\subsection{Postinflationary thermal fluctuations}
\label{subsec:Thermal_Fluctuations}
\vs{-5mm}
 
After inflation has ended and the Universe is reheated, the axion is essentially frozen out until the QCD phase transition. During that time the axion is subject to thermal fluctuations that drive it away from its value at the end of inflation. 

Since the potential is essentially flat, the motion along it can be treated as a random walk with $\Delta \theta \sim T/f_a$ per Hubble time. In the preinflationary scenario the axion is present from the very beginning of the thermal universe at the reheating temperature $T_\RH$. The maximum deviation until the QCD phase transition is then
	\begin{equation}
		 \sigma_\theta^{\rm thermal}
			\sim 
				\frac{T_\RH}{f_a}			
			\; .
	\end{equation}
The influence of the thermal fluctuations on the initial misalignment is similar to the one by quantum fluctuations and can be captured by the replacement $\theta_1^2~\rightarrow~\theta_1^2~+~(\sigma^{\rm thermal}_{\theta})^2$.

In the preinflationary scenario these fluctuations are suppressed, so they do not play a role for $\theta_1 \sim \mathcal{O}(1)$. However, they can become important for $\theta_1 \ll 1$ and can even dominate for certain values of the parameters. The same is true for the inflationary quantum fluctuations. We will later come back to this interesting scenario when we study a particular model.

\section{Early Relaxation using Inflation}
\label{sec:Early_Relaxation_using_Inflation}
\vs{-5mm}

\subsection{General idea}
\label{sec:General_Idea}
\vs{-5mm}

As already stated, the upper boundary of the axion window at $f_a \sim 10^{12} \GeV$ is based on the initial misalignment angle $\theta_1$ being $\mathcal{O}(1)$. However, smaller values of $\theta_1$ emerge naturally when the axion is able to relax to its minimum in the early Universe. Thus, the upper bound can be shifted to $f_a \gg 10^{12} \GeV$.

In this paper we follow the proposal in \cite{Dvali:1995ce} and use inflation to generate a temporal phase of strong QCD. During inflation QCD becomes strong when
	\begin{equation}
		\Lambda_\QCD
			\gtrsim 
				H_\Irm
			\; .
	\label{Eq:StrongQCD}
	\end{equation} 
The strong QCD results in the axion developing its potential and relaxing to its minimum. The efficiency of this relaxation, however, depends on the axion mass $m_a$, the Hubble parameter during inflation, and the number of $e$-folds.

For instance, if $m_a \gtrsim H_\Irm$ the Hubble friction is overcome and the relaxation to the minimum is very efficient. The crucial feature in this case is that the axion does not develop any isocurvature perturbations since those arise only for massless particles relative to the inflationary Hubble parameter. 

The relaxation is also possible for $m_a \lesssim H$ if the phase of strong QCD lasts long enough. In particular, since the axions dynamics in that case are given by
	\begin{equation}
		\theta(t_f)
			\sim 
				\theta(t_i)
				\exp
				\left(
					- \mathcal{N} 
						\frac{m_a^2}{3 H^2_\Irm}
				\right)
			\;, 
	\label{Eq:AxionSolEFold}
	\end{equation}
where $\theta(t_i) \sim \mathcal{O}(1)$ denotes the amplitude and $\mathcal{N}$ the number of $e$-folds, the initial phase of strong QCD must last for  
	\begin{equation}
		\mathcal{N} 
			\gg 
				\frac{3H^2_\Irm}{m^2_a}
			\; 
	\end{equation}
for the relaxation to be efficient.
	
Either way the potential vanishes at the end of this early phase of strong QCD and the axion remains frozen out very close to its minimum. When the ``ordinary" QCD phase transition activates the axion potential again, the resulting amplitude of the oscillations is naturally small, i.e., $\theta_1 << 1$.

\subsection{The QCD scale during inflation}
\vs{-5mm}

With QCD becoming strong during inflation when (\ref{Eq:StrongQCD}) is satisfied, it makes the impression that early relaxation is only applicable for small values of $H_\Irm$. This impression, however, is based on the assumption that parameters such as $m_a$ or $\Lambda_\QCD$ are unaltered during inflation. In fact, this assumption is not as natural as it seems.

For instance, consider the situation where the strong coupling constant $\alpha_s$ is fixed by VEV(s) of some scalar field(s) as it is common in generic supergravity and superstring theories. The minima of those scalars during inflation, in general, do not coincide with their low temperature minima due to both, thermal and quantum fluctuations, but also a displacement of the minima themselves. This could easily result in $\alpha_s \gg 1$ during inflation, thus making QCD strong with a $\Lambda_\QCD$ different from its late time value \cite{Dvali:1995ce}.

The same effect of displaced scalar VEVs during inflation can also change $\Lambda_\QCD$ in an indirect way. This is due to the VEV of the Standard Model Higgs $H$ being displaced as well. 

For example, consider the case of massive chaotic inflation, i.e., $V(\Phi) = m^2 \Phi^2/2$, together with the Higgs-portal coupling \cite{Kamada:2014ufa}
	\begin{equation}
		\Delta V
			=
				- \frac{1}{2}
				\kappa
				\Phi^2
				H^\dagger H
			\; ,
	\end{equation}
where $\kappa$ is some positive constant. From the Friedmann equations in the slow-roll approximation we have $3 H^2_\Irm M_\Prm^2 = m^2 \Phi^2/2$ so that the potential becomes
	\begin{equation}
		\Delta V
			=
				- \frac{1}{2}
				\kappa
				\left(
					6 \frac{H^2_\Irm}{m^2}
				\right)
				H^\dagger H
			\; .
	\end{equation}
We see that the Higgs receives a curvature of order $H_\Inf$ which shifts its VEV. Note that $\kappa \lesssim 10^{-6}$ for quantum corrections to be negligible in this particular example \cite{Kamada:2014ufa}.

Alternatively, a shift also appears when higher dimensional operators of the inflaton and the Higgs are included, e.g.,
	\begin{equation}
		\Delta V
			=	
				\frac{\lambda}{\Lambda^2}
				\Phi^4 H^\dagger H
			\; ,
	\end{equation}
where $\lambda < 1$ is a coupling constant and $\Lambda$ is some UV cutoff, say $M_\Prm$. In large field inflation the Higgs takes value of order of $M_\Prm$, such that the Higgs VEV itself becomes of order of $\lambda^{1/2} M_\Prm$.

Motivated by those examples, the Higgs VEV can be much larger during inflation. Since it contributes into the masses of the quarks, those will receive large shifts as well. This changes the running of $\alpha_s$ and thus $\Lambda_\QCD$. As can be read of from (\ref{Eq:AxionMassInstanonThermal}), the changed value of $\Lambda_\QCD$ automatically results in a changed value of $m_a$. For this reason, we will from now on denote parameters during inflation with the label ``$\Inf$".

Even though we will focus on the implications from the changed value $\Lambda_\QCD$ in the early Universe  on the axion, it should be noted that this could have interesting consequences on other theories. For example, this would allow for a production of primordial black holes from the confinement of our QCD as discussed in \cite{Dvali:2021byy}.

Let us now turn to the influence of inflation on the QCD scale $\Lambda_\QCD$ in the KSVZ and the DFSZ models. Both models consist of the Standard Model plus the following fields necessary to implement the PQ scenario:
\begin{itemize}
	\item KSVZ: A vectorlike fermion $Q \sim Q_\Lrm + Q_\Rrm$ in the SM representation (3,1,0), i.e., fundamental of SU(3)$_\crm$, singlet under SU(2)$_\Lrm$, and neutral under U(1)$_\Yrm$. Additionally, a SM-singlet complex scalar $S \sim (1,1,0)$.
	\item DFSZ: A second Higgs doublet $H_u \sim (1,2,- 1/2)$ [the usual Higgs doublet being $H_d \sim (1,2,+ 1/2)$] and a SM-singlet complex scalar $S \sim (1,1,0)$.
\end{itemize}
Since in these models the strong coupling constant is not controlled by the VEV of a moduli field, the shift comes purely from the displaced Higgs VEV. 

Recall that the first order $\beta$ function of QCD in dimensional regularization is given by
	\begin{equation}
		\beta
			\equiv
		\mu \frac{\drm \alpha_s}{\drm \mu}
			=
				- \frac{\alpha_s^2}{2\pi} b
			\; ,
		\label{Eq:BetaFunction}
	\end{equation}
where
	\begin{equation}
		b
			=
				11 - \frac{2}{3} n_f
			\; .
	\end{equation}
Here, $n_f$ denotes the number of active quarks in the considered energy range.

In the late Universe, given the measured value $\alpha_s(M_Z) = 0.1179$ \cite{PDG2020}, the ordinary differential equation (ODE) can be solved in the energy range $m_b \leq E \leq m_t$ where $n_f = 6$. Taking the value of the solution at $m_b$ as the new initial condition, the ODE can be solved in the energy range $m_c \leq E \leq m_b$ where $n_f = 5$, and so on. The QCD scale is identified with the scale where $\alpha_s = 1$ and is $\Lambda_\QCD \sim 0.28 \GeV$ at first loop.

Note that for the minimal KSVZ model, a heavy quark needs to be included in the running. Since this heavy quark is getting its mass from a corresponding singlet $S$, its mass is assumed to be roughly of order of $f_a$. This has an influence on the running only above $f_a$ and thus plays no role in the determination of the QCD scale. 
	
During inflation the quark masses are much larger as a result of the shifted Higgs VEV. These large masses will lead to the quarks being integrated out at much higher energies. Since the running of $\alpha_s$ becomes steeper with less active quarks, the emerging Landau pole is located at a much higher energy than with the smaller quark masses in the late Universe.  

In order to define the quark masses during inflation, we take into account the running of the Yukawa couplings in addition to the changed value of the Higgs VEV. Usually, the mass of the (heavy) quarks is defined at the corresponding mass scale in the $\overline{\rm MS}$ scheme, i.e., $m_q \sim v_H y_q(m_q)$, where $v_H$ is the Higgs VEV and $y_q$ is the corresponding Yukawa coupling. We take this as the defining relation and solve it numerically by making the approximation $y_q^{\rm \Inf} \sim y_q^{\rm today}$. Taking into account the Yukawa coupling in this way gives a slightly more precise definition of the quark masses than only using the shifted Higgs VEV, but it turns out that the final influence on the QCD scale is negligible.

Moreover, in the ordinary low energy situation the measured value of $\alpha_s(M_Z)$ served as the initial condition for the running. During inflation this condition is no longer applicable due to the changed quark masses. Therefore, we need to use as the initial condition the value of $\alpha_s$ at a higher scale that is unaffected by the larger quark masses, say, the GUT scale $\Lambda_{\rm GUT} \sim 10^{16} \GeV$ or the Planck scale $M_\Prm$. 

Since the value of $\alpha_s$ at such a large scale crucially depends on the field content, the particular value is very model dependent. In the DFSZ model without additional fields, the running is purely controlled by the known quarks and we have $\alpha_s(\Lambda_{\rm GUT}) \sim 0.0224$ or $\alpha_s(M_\Prm) \sim 0.0200$. For the KSVZ model, the heavy quark can have an influence if its mass is smaller than the energy at which the new initial condition is defined. Taking its mass to be not too far off from the GUT scale, $m_Q \sim \Lambda_{\rm GUT}$, its influence is rather small and the initial conditions of both the DFSZ and the KSVZ models coincide. It should be noted that the KSVZ singlet could be coupled to the inflaton itself so that the mass $m_Q$ is also shifted during inflation.

\begin{figure}[t]
	\includegraphics[scale=0.67,angle=0]{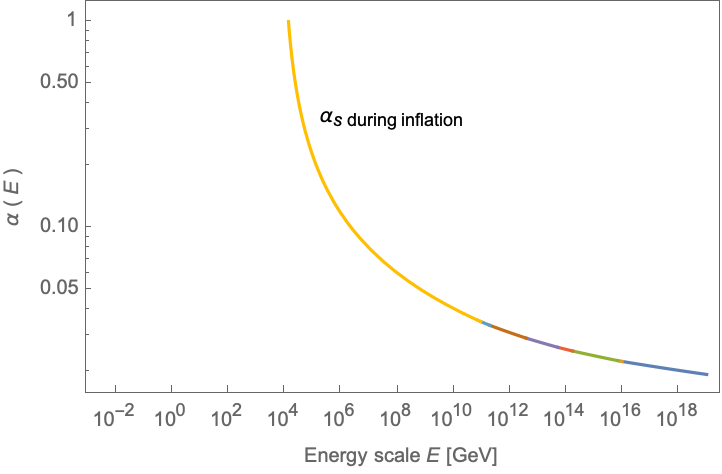}
	\caption{
		Running of $\alpha_s$ during inflation in the KSVZ model with the inflationary Higgs VEV chosen to be $v_H^\Inf \sim M_\Prm$ and the initial condition $\alpha_s^\Inf(M_\Prm) = \alpha_s(M_\Prm)$. The different colors indicate the parts with a different number $n_f$ of active quarks. We see that the QCD scale is much larger compared to its value in the late Universe.}
	\label{Fig:RunningA}
\end{figure}

In order to calculate the QCD scale, we solve (\ref{Eq:BetaFunction}) in the same manner as in the late Universe but incorporate the changes we just discussed. For illustration, let us use the following set of parameters in the KSVZ model. First, the mass of the heavy KSVZ quark is fixed at $m_Q \sim \Lambda_{\rm GUT}$. Next, the late time value of $\alpha_s$ at $M_\Prm$ is used as the initial condition. Lastly, the value of the inflationary Higgs VEV is chosen, i.e., $v_H^\Inf \sim M_\Prm$. We depict the resulting running of $\alpha_s$ in Fig. \ref{Fig:RunningA}. As can be read of, with these parameters the QCD scale is located at $\Lambda^\Inf_\QCD \sim 10^5 \GeV$. 

\begin{figure}[t]
	\includegraphics[scale=0.67,angle=0]{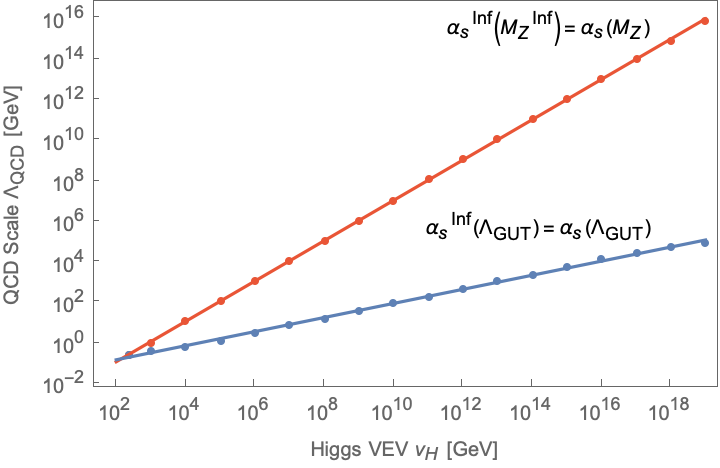}
	\caption{
	Values of the QCD scale for various values of the inflationary Higgs VEV $v_H^\Inf$ and the corresponding linear fit. The red points have $\alpha_s^\Inf(m_Z^\Inf) = \alpha_s(m_Z)$ as an initial condition for the running while the blue points have $\alpha_s^\Inf(\Lambda_{\rm GUT}) = \alpha_s(\Lambda_{\rm GUT})$. In both, the KSVZ quark mass was fixed to be of order of $f_a \sim \Lambda_{\rm GUT}$.}	
	\label{Fig:QCDScales}
\end{figure}

We repeat this procedure for various values of the inflationary Higgs VEV. The resulting values of the QCD scales are plotted in blue in Fig. \ref{Fig:QCDScales} and can be fitted by
	\begin{equation}
		\Lambda_\QCD^\Inf (v_H)
			\sim 
				10^5 \GeV
				\left(
					\frac{v_H}{M_\Prm}
				\right)^{5/14}
			\; .
	\label{Eq:LambdaInf1}
	\end{equation}
Our analyses shows that for the given choice of parameters and initial condition the QCD scale during inflation is significantly changed. This effect has also been found in \cite{Matsui:2020wfx}. The different exponent in their calculation comes from including the running of Yukawa couplings. 

The analysis can be repeated for the DFSZ model. By comparing the running of both models at hand with the fixed mass of the KSVZ quark $m_Q \sim f_a$, we find that for the parameter space of our interest, i.e., $f_a > 10^{12}$, the KSVZ quark has negligible influence on the running of $\alpha_s$. This conclusion is not changed when a coupling between the inflaton and the KSVZ singlet is included so that $f_a$ gets shifted during inflation. In particular, the influence on the running gets even more negligible.

In order to demonstrate the impact of the initial condition on the value of the QCD scale, we solve the ODE with an alternative initial condition, say $\alpha_s^\Inf(m_Z^\Inf) = \alpha_s(m_Z)$. Of course, additional fields are required for this but let us not delve into this and just calculate the running to illustrate the influence of the initial condition. The resulting values are plotted in red in Fig. \ref{Fig:QCDScales} and can be fitted by
	\begin{equation}
		\Lambda_\QCD^\Inf (v_H)
			\sim 
				10^{15} \GeV
				\left(
					\frac{v_H}{M_\Prm}
				\right)
		\; .
	\label{Eq:LambdaInf2}
	\end{equation}
From this expression it follows that the QCD scale can, in principle, take even larger values just by running.

\subsection{Location of minima during inflation}
\label{sec:Shifted_Minima_of_DFSZ_and_KSVZ}
\vs{-5mm}

While the idea of early relaxation is surprisingly simple, a particular realization needs to take into account that the minima of the axion during both phases of strong QCD must coincide. Indeed, if this was not the case, the axion would relax to its early time minimum but when the ordinary QCD phase transition takes place, the axion is not necessarily located close to its true minimum. Fine-tuning would be required, which renders the proposed mechanism not better than simply fine-tuning $\theta_1 \ll 1$. 

For instance, this is the case when the axion receives a large temporal additional mass in the early Universe. The crucial word here is additional. The axion then relaxes to its minimum at that time, but the minimum is, in general, very different from the true minimum at late times. This reintroduces the requirement of fine-tuning.

Thus, in order to achieve the coincidence of the minima, it is necessary that the axion's potential in the early Universe arises in the same way as its potential in the late Universe -- by nonperturbative QCD effects or, in other words, by making QCD strong. This however is not enough since the physics that give rise to this early phase of strong QCD can also change the location of the minimum. Thus, the necessary condition is that the early phase of QCD emerges in such a way that the axions minimum during that phase coincides sufficiently with the minimum of the late Universe.

In the following we calculate the displacement of the minima for the KSVZ and the DFSZ axions during inflation. We perform the calculation with all parameters taken to be different during inflation and for a first approximation assume a step-function-like transition from the inflationary values to the later ones. For a clearer depiction, we introduce the label ``$0$" for a value in the late Universe in addition to the label ``$\Inf$" for a value during inflation.

The location of the minima is seen most conveniently by considering the axion Lagrangian in the presence of the $\theta$-term,
	\begin{equation}
		\mathcal{L}
			\supset
				\left(
					\bar{\theta}
					+ \frac{\tilde{a}}{f_a}
				\right)
				\frac{\alpha_s^2}{8\pi}
				G^a_{\mu\nu} \tilde{G}^{a \, \mu \nu}
			\; ,
	\end{equation}
where $\bar{\theta}$ is the $CP$ violating parameter with the contribution from QCD and the quark masses, i.e.,
	\begin{equation}
		\bar{\theta}
			=
				\theta_\QCD + \arg \det y_u y_d
			\; .
	\end{equation}
After QCD gets strong, this gives rise to a potential $V(\theta~+~\tilde{a}/f_a)$, which according to the Vafa-Witten theorem has its minimum at $\langle \tilde{a} \rangle = - \bar{\theta} f_a$. The physical axion field is then identified as
	\begin{equation}
		a 
			= 
				\tilde{a} 
				- \langle \tilde{a} \rangle
			\; .
	\end{equation}
Because of different values of the parameters during the cosmological evolution, the minima do not coincide and thus the physical axion field itself is also different. Regarding the original field $\tilde{a}$, however, it depends on the particular axion model. 

In the case of the KSVZ, the definition of $\tilde{a}$ does not depend on any additional parameter so that it is the same at any time. Hence, we can write down the axion VEVs in the late Universe and during inflation as
	\begin{align}
		\langle \tilde{a} \rangle_0
			&=
				- \bar{\theta}_0 f^a_0
			\; , 
			\label{Eq:AxionVEV0}\\
		\langle \tilde{a} \rangle_\Inf 
			&=
				- \bar{\theta}_\Inf f^a_\Inf
			\; .
			\label{Eq:AxionVEVInf}
	\end{align}
and using these, the resulting displacement of the vacua $\Delta \theta$ is given by
	\begin{equation}
		\Delta \theta_{\rm KSVZ}
			\equiv
				\left|
					\frac{\langle \tilde{a} \rangle_\Inf
					- \langle \tilde{a} \rangle_0}{f^a_0} 
				\right|
			=
				\left|
					\bar{\theta}_0
					- \bar{\theta}_\Inf \frac{ f^a_\Inf}{f^a_0}
				\right|
			\; .	
		\label{Eq:DisplacementKSVZ}			
	\end{equation}

Let us now turn to the DFSZ axion. Here the situation is more involved due to the original axion field $\tilde{a}$ depending on the VEVs of the two Higgs doublets $H_u$ and $H_d$ and the singlet $S$. For concreteness, let us take the DFSZ-I model where the Yukawa sector is given by
	\begin{equation}
		\mathcal{L}
			\supset
				- y_u H_u \bar{Q}_L u_R
				- y_d H_d \bar{Q}_L d_R
				- y_e H_d \bar{E}_L e_R
				+ \hrm.\crm.
			\; .
	\end{equation}
Given a proper scalar potential, all three scalar fields develop a VEV,
	\begin{align}
		H_u
			\supset
				\frac{v_u}{\sqrt{2}}
				\erm^{i \frac{a_u}{v_u}}
				\begin{pmatrix}
					1 \\
					0
				\end{pmatrix}
			\; ,
		H_d
			\supset
				\frac{v_d}{\sqrt{2}}
				\erm^{i \frac{a_d}{v_d}}
				\begin{pmatrix}
				0 \\
				1
				\end{pmatrix}
			\; ,	
		S 
			\supset
				\frac{v_s}{\sqrt{2}}
				\erm^{i \frac{a_s}{v_s}}
			\; ,
	\end{align}
where any modes not containing the axion are neglected. Each Goldstone transforms under a PQ transformation as $a_i \rightarrow a_i + \kappa \chi_i v_i$ where $\chi_i$ denote the PQ charges of the corresponding scalar. The corresponding PQ current is
	\begin{align}\nonumber
		J_\mu^\PQ 
		\Big|_a
			&\supset
				- \chi_S
					S^\dagger 
					i \overset{\leftrightarrow}{\partial^\mu}
					S
				- \chi_u
					H_u^\dagger 
					i \overset{\leftrightarrow}{\partial^\mu}
					H_u
				- \chi_d
					H_d^\dagger 
					i \overset{\leftrightarrow}{\partial^\mu}
					H_d
					\\
			&= 
				\sum_{i=S,u,d}
				\chi_i
				v_i \partial_\mu a_i
			\; .
	\end{align} 
The axion field is then defined as 
	\begin{equation}
		\tilde{a}
			=
				\frac{1}{v_a}
				\sum_{i=S,u,d} \chi_i v_i a_i
			\; ,
			\qquad
		v_a^2 
			=
				\sum_{i=S,u,d} 
				\chi_i^2 v_i^2
			\; ,
		\label{Eq:DFSZAxion}
	\end{equation}
so that $J_\mu^\PQ |_a = v_a \partial_\mu \tilde{a}$ and under the PQ transformation the axion transforms as $\tilde{a} \rightarrow \tilde{a} + \kappa v_a$. 

The PQ charges are fixed by requiring the PQ invariance of the operator $H_u H_d S^{\dagger2}$ that couples the singlet to the doublets, implying $\chi_u +\chi_d - 2\chi_s = 0$ and the orthogonality between $J_\mu^\PQ |_a$ and the hypercharge current $J_\mu^Y|_a = \sum_i Y_i v_i \partial_\mu a_i$ implying $-\chi_u v_u^2 +\chi_d v_d^2 = 0$. Choosing the same overall normalization as in \cite{DiLuzioAxionLandscape}, the charges are
	\begin{align}\nonumber
		\chi_S &= 1 \; , 	&
		\chi_u &= 2 \left( \frac{v_d}{v} \right)^2 \; , 	&
		\chi_d &= 2 \left( \frac{v_u}{v} \right)^2 \; ,		\\
		&	&	
		&\equiv 2 \cos^2 \beta 	&	
		&\equiv 2 \sin^2 \beta
	\label{Eq:PQCharges}
	\end{align}
where we have introduced the mixing angle between the two doublets, $\beta$, and $v^2 \equiv v_u^2 + v_d^2 = (246 \GeV)^2$ is the electroweak VEV.  

In addition, it turns out to be helpful to introduce the field 
	\begin{equation}
		a_D
			=
				\frac{1}{v_D}
				\sum_{i=u,d} \chi_i v_i a_i
			\; ,
			\qquad
		v_D^2 
			=
				\sum_{i=u,d} 
				\chi_i^2 v_i^2
			=
				v^2 \sin^2 2 \beta
			\; ,
	\end{equation}
which is orthogonal to the ``Hypercharge Goldstone" in the plane spanned by $a_u$ and $a_d$. Using this definition and the PQ charges from (\ref{Eq:PQCharges}), the DFSZ-I axion in (\ref{Eq:DFSZAxion}) can be written as
	\begin{equation}
		\tilde{a}
			=
				\frac{1}{v_a}
				(v_D a_D
				+ v_s a_s)
			\; ,
			\qquad 
		v_a^2 
			=
				v_D^2 + v_S^2
			\; .
	\end{equation}	
Written in this way, the DFSZ axion is a superposition of two and not three fields, which is helpful for visualization.

Even though Eqs.~(\ref{Eq:AxionVEV0}) and (\ref{Eq:AxionVEVInf}) are also valid for the DFSZ, calculating the displacement by just subtracting the VEVs from each other as in the KSVZ case does not give the correct initial amplitude for today's axion $\tilde{a}_0$. The reason is that the axion potential in the late Universe emerges along $\tilde{a}_0$. Hence, it is the projection on $\tilde{a}_0$ that results in the correct amplitude for the late time oscillations,
	\begin{align} \nonumber
		\Delta \theta_{\rm DFSZ}
			&\equiv
				\left|
					\frac{\langle \tilde{a} \rangle_\Inf |_{\tilde{a}_0}
					- \langle \tilde{a} \rangle_0}{f^a_0} 
				\right|
			\\
			&= 
				\left|
					\bar{\theta}^0
					- \bar{\theta}^\Inf \frac{f_a^\Inf}{f_a^0}
					P
				\right|
			\; .
	\label{Eq:DisplacementDFSZ}			
	\end{align}
Here, we used the correct normalization for the axion decay constant, i.e., $f_a = v_a/2N$ with $N=3$ for the DFSZ model, and introduced the projection factor $P$,
	\begin{align}\nonumber
		P
			=
				&\left(
					\frac{\chi_u^\Inf v_u^\Inf}{v_a^\Inf}
				\right)
				\left(
					\frac{\chi_u^0 v_u^0}{v_a^0}
				\right)
				+ \left(
					\frac{\chi_d^\Inf v_d^\Inf}{v_a^\Inf}
				\right)
				\left(
					\frac{\chi_d^0 v_d^0}{v_a^0}
				\right) \\ \nonumber
				&+ \left(
					\frac{v_S^\Inf}{v_a^\Inf}
				\right)
				\left(
					\frac{v_S^0}{v_a^0}
				\right) \\
				\sim
				&\left(
					\sin \beta^0
					+ \cos \beta^0 
				\right)
				\sin 2\beta^0
				\frac{v^\Inf}{v_a^\Inf}
				\frac{v^0}{v_a^0}
				+ \left(
					\frac{v_S^\Inf}{v_a^\Inf}
				\right)
			\; ,
	\end{align}
where in the second row we used $v_S^0 \sim v_a^0$ for today's axion to be invisible and, in addition, we assumed both doublets to get similar shifting from the inflaton so that $\beta^\Inf \sim \pi/4$. The first term is strongly suppressed compared to the second one due to the factor $v^0/v_a^0 \ll 1$ and the fact that $\beta^0$ cannot be too large for the perturbativity of the top and bottom Yukawa couplings \cite{Bjorkeroth:2019jtx}. Hence, the difference between the KSVZ and DFSZ models comes down to the factor of $P \sim v_s^\Inf/v_a^\Inf \lesssim 1$. 

Within the KSVZ and the DFSZ models $\bar{\theta}$ is, per se, a free parameter but let us assume it takes its natural value following from its renormalization \cite{ELLIS1979141},
	\begin{equation}
		\bar{\theta}
			\sim
				\left(
					\frac{\alpha}{\pi}
				\right)^2
				s_1^2 s_2 s_3 \sin \delta \,
				\frac{m_s^2 m_c^2}{m_W^2}
			\sim
				10^{-16}
			\; .
	\label{Eq:ThetaRenorm}
	\end{equation}
Here $m_s$ is the mass of the strange quark, $m_c$ is the mass of the charm quark, and $m_W$ is the mass of the $W$ boson. Furthermore, the $s_k \equiv \sin \phi_k$ are the mixing angles and $\delta$ is the $CP$ odd phase of the CKM matrix in the original parametrization of Kobayashi and Maskawa. Evaluated during inflation, the running is not that much altered since the factor including the masses is independent of the Higgs VEV, the running of the couplings is very slow, and the parameters of the CKM matrix are unchanged. Hence, it is reasonable to take $\bar{\theta}_\Inf \sim \bar{\theta}_0 \sim 10^{-16}$.

As a result, the initial misalignment angle $\theta_1 \sim \Delta \theta$ can become extremely small for both, the KFSZ and the DFSZ models, if the axion efficiently relaxes to the minimum. In particular, if the only contribution to the axion energy density came from $\theta_1$, it would be so small that any value of $f_a^0$ would result in the axion making up a negligible amount of the dark matter. 

\subsection{The axion window}
\label{sec:Thermal_Fluctuations}
\vs{-5mm}

Since the the axion's inflationary vacuum essentially coincides with its late time vacuum for the KSVZ and the DFSZ models, the value of $\theta_1$ is determined by the efficiency of the relaxation. Thus, for sufficiently long inflation according to (\ref{Eq:AxionSolEFold}), the region with $H_\Irm \lesssim \Lambda_\QCD^\Inf$ becomes viable. Without further field content than the one necessary for the KSVZ and the DFSZ models, the maximum inflationary QCD scale is $\Lambda_\QCD^\Inf \sim 10^5 \GeV$.

An interesting scenario is given when the relaxation is extremely efficient so that the contribution from the misalignment angle itself becomes completely negligible. Including all fluctuations, the final axion abundance is given by
	\begin{equation}
		\frac{\Omega_a }{\Omega_{\rm CDM}}
			\sim
				\left(
					\frac{f_a}{10^{12} \GeV}
				\right)^{7/6} 
				\bigg[
					\theta_1^2
					+ \sigma_\theta^2
					+ \Big( \sigma_\theta^{\rm thermal} \Big)^2
				\bigg]
				\; .	
	\label{Eq:TotalTheta}
	\end{equation}
With negligible $\theta_1$, the dark matter density will be dominated by quantum or thermal fluctuations, which in turn depend on the reheating temperature $T_\RH$ and on the inflationary Hubble scale $H_\Irm$. In the case when either of the types of fluctuations dominates, we can solve (\ref{Eq:TotalTheta}) to find the parameter space where the axion makes up all the dark matter, namely,
	\begin{align}
		H_\Irm
			&\sim
				10^{13} \GeV
				\left(
					\frac{f_a}{10^{12} \GeV}
				\right)^{5/6}
				:
			& & H_\Irm \gg T_\RH \\	
		T_\RH 
			&\sim	
				10^{12} \GeV
				\left(
					\frac{f_a}{10^{12} \GeV}
				\right)^{5/6}
				:
			& & H_\Irm \ll T_\RH
			\; .
	\label{Eq:DMRatioExtrem}
	\end{align}

When quantum fluctuations dominate over the postinflationary thermal fluctuations, it is not possible to reach values for $f_a$ that are much larger than $10^{12}\,\GeV$. This is due to the upper bound on $H_\Irm$ from inflationary gravitational waves as dictated by (\ref{Eq:GravWaveBound}). Hence, for such an efficient relaxation the cosmological bound can only be avoided by large reheating temperatures. In particular, using the maximum thermalization temperature defined in (\ref{Eq:TMax}) as an upper bound for the reheating temperature allows us to give a constraint on $f_a$ in terms of $H_\Irm$,
	\begin{equation}
		f_a	
			\lesssim	
				10^{15} \GeV
				\left(
					\frac{\epsilon_\eff}{1}
				\right)^{6/5}
				\left(
					\frac{H_\Irm}{10^{11} \GeV}
				\right)^{3/5}
			\; .
		\label{Eq:EffRelax1}
	\end{equation}
	
	\begin{figure}[t]
		\includegraphics[scale=0.67,angle=0]{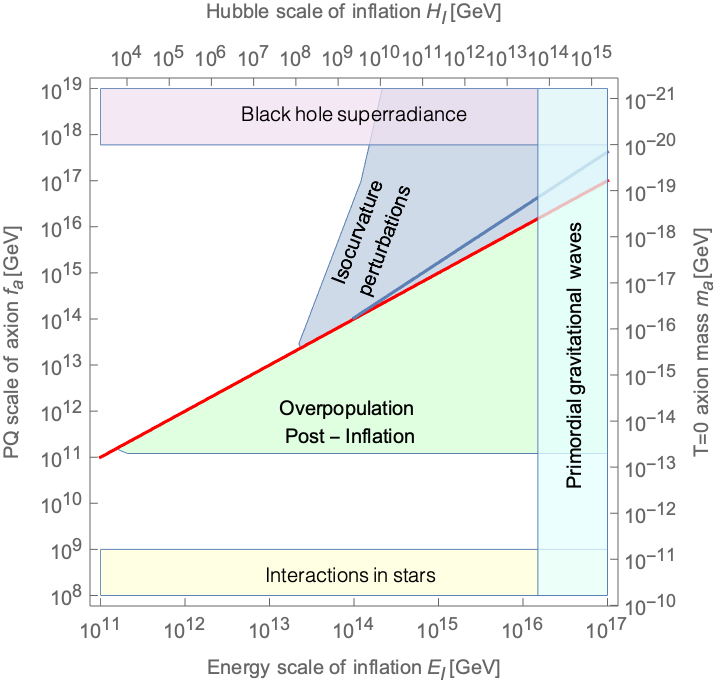}
		\includegraphics[scale=0.67,angle=0]{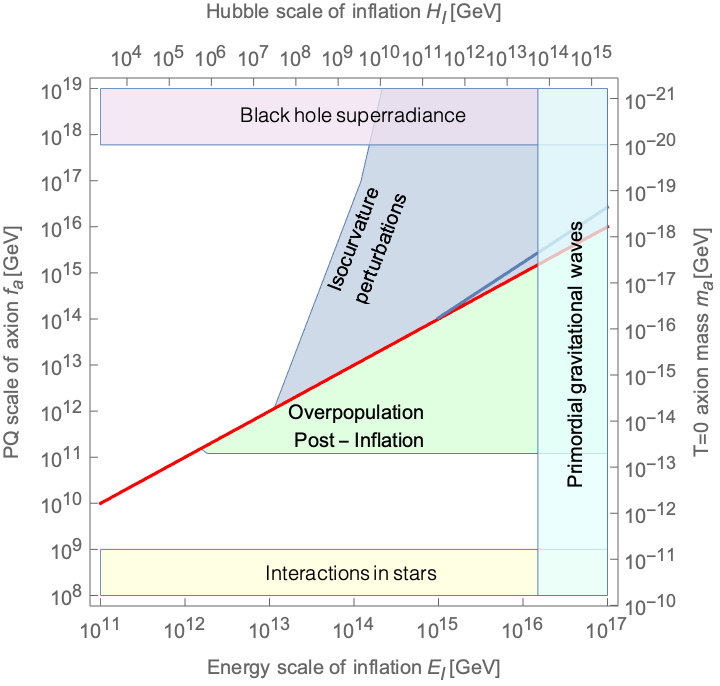}
		\caption{Constraints on the axion for very efficient reheating, $\epsilon_\eff \sim 1$ (upper) and $\epsilon_\eff \sim 10^{-1}$ (lower). The blue line depicts the upper bound for our minimal scenarios. Hence, the region between the red line and the blue line defines the parameter space where the axion dynamically relaxes to the minimum during inflation and later makes up all the dark matter.}
		\label{Fig:HubbletoPQScale}
	\end{figure}

We depict the maximum of this bound by a blue line in Fig. \ref{Fig:HubbletoPQScale} for $\epsilon_\eff \sim 1$ and for $\epsilon_\eff \sim 10^{-1}$. The allowed parameter space is thus defined by the region enclosed by the blue and red lines. For $\epsilon_\eff \sim 10^{-2}$ there is already no more parameter space left. Overall we can say, that efficient relaxation of $\theta_1$ results in a very narrow allowed region. 

On top of that, this region lies in the isocurvature constrained part of the parameter space. Hence, it is only allowed if the axion avoids the isocurvature constrain for which there are several known possibilities (see \cite{DiLuzioAxionLandscape} for an overview). One of those possibilities is given by the fact that if the axion acquires a sufficiently high mass the constrain is relaxed \cite{Jeong:2013xta}. In this case, the QCD scale during inflation has to fulfill not only $H_\Irm \lesssim \Lambda_\QCD^\Inf$ in order to make QCD strong but the more restrictive condition 
	\begin{equation}
		H_\Irm
			\lesssim
				m_a^\Inf
			\sim
				\begin{cases}
					\frac{\sqrt{(\Lambda^\Inf_\QCD)^3 \, m^\Inf_u}}{f_a^\Inf} 
					&:m_u^\Inf \lesssim \Lambda_\QCD^\Inf
					\; ,
					\vspace{0.2cm}
					\\
					\frac{(\Lambda^\Inf_\QCD)^2}{f_a^\Inf}
					&:m_u^\Inf > \Lambda_\QCD^\Inf
					\; .
				\end{cases}				
	\label{Eq:EffRelax2}
	\end{equation}
Here we use different expressions for the axion mass depending on whether there are light quarks or not. In the situation where there are no light quarks, as described by the lower equation, we used the axion mass derived from the dilute instanton gas approximation.  

Alternatively, by plugging (\ref{Eq:EffRelax2}) into (\ref{Eq:EffRelax1}), we can derive a condition on $\Lambda_\QCD^\Inf$, 
	\begin{equation}
		\Lambda_\QCD^\Inf
			\gtrsim
				\begin{cases}
					10^{15} \GeV
					\left(
						\frac{f_a}{10^{16} \GeV}
					\right)^{16/9}
					\left(
						\frac{M_\Prm}{v_H^\Inf}
					\right)^{1/3}
					&\; ,
					\vspace{0.2cm}		
					\\		
					10^{14} \GeV
					\left(
						\frac{f_a}{10^{16} \GeV}
					\right)^{4/3}
					\left(
					\frac{1}{\epsilon_\eff}
					\right)
					&\; .
				\end{cases}
	\label{Eq:ConditionLambdaQCD}
	\end{equation}
This condition has the advantage that once $\Lambda_\QCD^\Inf(v_H^\Inf)$ is known, one can immediately say if the model is compatible with very efficient relaxation. 

In the KSVZ and DFSZ models, where $\Lambda_\QCD^\Inf(v_H^\Inf)$ is given by (\ref{Eq:LambdaInf1}), it is not possible to satisfy condition (\ref{Eq:ConditionLambdaQCD}) for any point in the viable parameter space. Without additional field content that modifies the running of the strong coupling, $\Lambda_\QCD^\Inf$ is simply not high enough. Thus, in these two models such an efficient relaxation must be avoided. For instance, this can be achieved by requiring $\theta_1 \gtrsim \sigma_{\theta}$. Using (\ref{Eq:AxionSolEFold}), this translates into
	\begin{equation}
		\mathcal{N}
			\lesssim
				\frac{3 H^2_\Irm}{(m_a^\Inf)^2}
				\ln \left(
					\frac{2 \pi f_a}{H_I}
				\right)
			\; .
	\end{equation} 
For $H_\Irm \sim \Lambda_\QCD^\Inf \sim 10^5 \GeV$ and $f_a \sim 10^{16} \GeV$, this bound gives $\mathcal{N} \lesssim 10^{22}$, which is not very restrictive.

Let us also consider for demonstration purpose the alternative running of $\Lambda_\QCD^\Inf(v_H^\Inf)$ given by (\ref{Eq:LambdaInf2}), which requires additional field content such that the initial condition of the running was $\alpha_s^\Inf (M_Z^\Inf) = \alpha_s(M_Z)$. At the point with $f_a^\Inf~\sim~10^{16}~\GeV$ and $H_\Irm~\sim~10^{13} \GeV$ in the allowed parameter space for $\epsilon_\eff \sim 1$, condition (\ref{Eq:ConditionLambdaQCD}) is satisfied for $v_H^\Inf \gtrsim 10^{17} \GeV$. Thus, it would be possible to incorporate efficient relaxation into such a model. 

\section{Summary and Outlook}
\label{sec:Summary_and_Outlook}
\vs{-5mm}

With the axion being one of the best motivated dark matter candidates and the booming axion experimental program, an understanding of the axions parameter space is of utter importance. In this work we studied the mechanism proposed in \cite{Dvali:1995ce} in the context of KSVZ and DFSZ axion models and show that there is a part of the parameter space that becomes viable. In particular, our analysis is automatically valid for the two-form implementation of the axion due to duality below the Peccei-Quinn scale and a minor influence of the KSVZ and DFSZ fields on the running.

When QCD becomes strong during inflation, i.e., when $H_\Irm \lesssim \Lambda_\QCD^\Inf$, the axions potential is activated prior to the time when the ordinary misalignment mechanism takes place. In this way it can relax to the minimum so that, when the ordinary QCD phase transition takes place, the axion is located close to its minimum. Hence, the initial misalignment angle naturally takes small values.

By means of a Higgs portal between the inflaton and our Higgs doublet or by higher order operators of those fields, the quark masses are much larger during inflation than they are in today's Universe. This changes the running of the strong coupling constant and thus can result in a larger value of the QCD scale. In particular, using $\alpha_s^\Inf (M_{\rm GUT}) = \alpha_s (M_{\rm GUT})$ as the initial condition for the running as dictated by the minimal field content of the KSVZ and DFSZ models, $\Lambda_\QCD^\Inf$ can be as large as $10^5 \GeV$. But using additional field content such that $\alpha_s^\Inf (M_Z^\Inf) = \alpha_s (M_Z)$ would be the initial condition, $\Lambda_\QCD^\Inf$ can be as large as $10^{15} \GeV$. 

For the minimal versions of the KSVZ and DFSZ models, the displacement of the minimum by the physics that make QCD strong during inflation is completely negligible. In addition, the axion develops its potential in the same way as in the late Universe, namely, by means of the same nonperturbative QCD effects. Hence, the relaxation is guarantied to suppress the initial misalignment angle where the amount of suppression depends on the duration of this early phase of strong QCD. 

In the case when $m_a^\Inf \gtrsim H_\Irm$ or inflation lasts long enough, the axion is diluted to an extend where the initial misalignment angle becomes completely negligible. The today's energy density of the axion is then dictated by inflationary quantum fluctuations and postinflationary thermal fluctuations. Because of the bound on the inflationary Hubble scale coming from inflationary tensor perturbations, only thermal fluctuations can result in the axion making up all the dark matter for values of $f_a \gtrsim 10^{12} \GeV$.  The amount of thermal fluctuations is controlled by the reheating temperature, which can be bounded by above by the maximum thermalization temperature defined in (\ref{Eq:TMax}). For very efficient reheating, $10^{-1} \lesssim \epsilon_\eff \lesssim 1$, this results in a relatively narrow region of applicability, while for less efficient reheating there is no viable parameter space. In particular, the point with $f_a \sim 10^{16} \GeV$ and $H_\Irm \sim 10^{13} \GeV$ lies in that allowed region, which is very interesting because the PQ scale is located at the GUT scale and the inflationary Hubble parameter is very close to the experimental constraint. 

Since this narrow region of applicability is furthermore constrained by isocurvature perturbations, these must be avoided in addition. Requiring for this purpose the axion mass $m_a^\Inf$ to be larger than the inflationary Hubble parameter $H_\Irm$ results in a constrain on $\Lambda_\QCD^\Inf$. Unfortunately, this constrain is not satisfied for the minimal KSVZ and DFSZ models where the modification of the QCD scale comes only from the displacement of the Higgs VEV. A modification of the running due to additional fields or a direct enhancement via a moduli field could alleviate this, however.
 
An interesting possibility for the realization of such a modification is given by the embedding of the PQ axion into GUTs such as SO(10) with an intermediate breaking into SU(5) or the Pati-Salam group. These gauge groups do not only modify the initial condition of the running, but their gauge couplings also have a steeper running towards the strong coupling. This could possibly be used during inflation to have a larger $\Lambda_\QCD^\Inf$, thus allowing for the implementation of a GUT or $M_\Prm$ scale axion without the need of moduli fields.  

\acknowledgments

I am  grateful to Gia Dvali for fruitful discussions, reading the manuscript, and useful comments. Furthermore, I am thankful for the Physikzentrum in Bad Honnef for its kind hospitality and great discussions I had there while working on the project. Finally, I want to thank the IMPRS research school of the Max-Planck Institute for Physics for the opportunity to participate in their Ph.D. program.

\setlength{\bibsep}{5pt}
\setstretch{1}
\bibliographystyle{utphys}
\bibliography{refs}

\providecommand{\href}[2]{#2}\begingroup\raggedright\begin{thebibliography}{10}

\bibitem{MisalignmentPRESKILL}
J.~Preskill, M.~B. Wise, and F.~Wilczek, ``Cosmology of the invisible axion,''
  \href{http://dx.doi.org/https://doi.org/10.1016/0370-2693(83)90637-8}{{\em
  Physics Letters B} {\bfseries 120} no.~1, (1983) 127--132}.
  \url{https://www.sciencedirect.com/science/article/pii/0370269383906378}.

\bibitem{MisalignmentDINE}
M.~Dine and W.~Fischler, ``The not-so-harmless axion,''
  \href{http://dx.doi.org/https://doi.org/10.1016/0370-2693(83)90639-1}{{\em
  Physics Letters B} {\bfseries 120} no.~1, (1983) 137--141}.
  \url{https://www.sciencedirect.com/science/article/pii/0370269383906391}.

\bibitem{MisalignmentAbbott}
L.~F. Abbott and P.~Sikivie, ``{A Cosmological Bound on the Invisible Axion},''
  \href{http://dx.doi.org/10.1016/0370-2693(83)90638-X}{{\em Phys. Lett. B}
  {\bfseries 120} (1983) 133--136}.

\bibitem{Dvali:1995ce}
G.~R. Dvali, ``{Removing the cosmological bound on the axion scale},''
  \href{http://arxiv.org/abs/hep-ph/9505253}{{\ttfamily arXiv:hep-ph/9505253}}.

\bibitem{Arvanitaki:2009fg}
A.~Arvanitaki, S.~Dimopoulos, S.~Dubovsky, N.~Kaloper, and J.~March-Russell,
  ``{String Axiverse},''
  \href{http://dx.doi.org/10.1103/PhysRevD.81.123530}{{\em Phys. Rev. D}
  {\bfseries 81} (2010) 123530},
  \href{http://arxiv.org/abs/0905.4720}{{\ttfamily arXiv:0905.4720 [hep-th]}}.

\bibitem{dvali2005threeform}
G.~Dvali, ``{Three-form gauging of axion symmetries and gravity},''
  \href{http://arxiv.org/abs/hep-th/0507215}{{\ttfamily arXiv:hep-th/0507215}}.

\bibitem{PhysRevD.105.085020}
O.~Sakhelashvili, ``Consistency of the dual formulation of axion solutions to
  the strong $cp$ problem,''
  \href{http://dx.doi.org/10.1103/PhysRevD.105.085020}{{\em Phys. Rev. D}
  {\bfseries 105} (Apr, 2022) 085020}.
  \url{https://link.aps.org/doi/10.1103/PhysRevD.105.085020}.

\bibitem{Dvali:2022fdv}
G.~Dvali, ``{Strong-$CP$ with and without gravity},''
  \href{http://arxiv.org/abs/2209.14219}{{\ttfamily arXiv:2209.14219
  [hep-ph]}}.

\bibitem{Dvali:2018dce}
G.~Dvali, C.~Gomez, and S.~Zell, ``{A Proof of the Axion?},''
  \href{http://arxiv.org/abs/1811.03079}{{\ttfamily arXiv:1811.03079
  [hep-th]}}.

\bibitem{KSVZ1}
J.~E. Kim, ``{Weak Interaction Singlet and Strong CP Invariance},''
  \href{http://dx.doi.org/10.1103/PhysRevLett.43.103}{{\em Phys. Rev. Lett.}
  {\bfseries 43} (1979) 103}.

\bibitem{KSVZ2}
M.~Shifman, A.~Vainshtein, and V.~Zakharov, ``Can confinement ensure natural cp
  invariance of strong interactions?''
  \href{http://dx.doi.org/https://doi.org/10.1016/0550-3213(80)90209-6}{{\em
  Nuclear Physics B} {\bfseries 166} no.~3, (1980) 493--506}.
  \url{https://www.sciencedirect.com/science/article/pii/0550321380902096}.

\bibitem{DFSZ1}
A.~R. Zhitnitsky, ``{On Possible Suppression of the Axion Hadron Interactions.
  (In Russian)},'' {\em Sov. J. Nucl. Phys.} {\bfseries 31} (1980) 260.

\bibitem{DFSZ2}
M.~Dine, W.~Fischler, and M.~Srednicki, ``A simple solution to the strong cp
  problem with a harmless axion,''
  \href{http://dx.doi.org/https://doi.org/10.1016/0370-2693(81)90590-6}{{\em
  Physics Letters B} {\bfseries 104} no.~3, (1981) 199--202}.
  \url{https://www.sciencedirect.com/science/article/pii/0370269381905906}.

\bibitem{DiLuzio:2017pfr}
L.~Di~Luzio, F.~Mescia, and E.~Nardi, ``{Window for preferred axion models},''
  \href{http://dx.doi.org/10.1103/PhysRevD.96.075003}{{\em Phys. Rev. D}
  {\bfseries 96} no.~7, (2017) 075003},
  \href{http://arxiv.org/abs/1705.05370}{{\ttfamily arXiv:1705.05370
  [hep-ph]}}.

\bibitem{Plakkot:2021xyx}
V.~Plakkot and S.~Hoof, ``{Anomaly ratio distributions of hadronic axion models
  with multiple heavy quarks},''
  \href{http://dx.doi.org/10.1103/PhysRevD.104.075017}{{\em Phys. Rev. D}
  {\bfseries 104} no.~7, (2021) 075017},
  \href{http://arxiv.org/abs/2107.12378}{{\ttfamily arXiv:2107.12378
  [hep-ph]}}.

\bibitem{Diehl:2023uui}
J.~Diehl and E.~Koutsangelas, ``{DFSZ-Type Axions and Where to Find Them},''
  \href{http://arxiv.org/abs/2302.04667}{{\ttfamily arXiv:2302.04667
  [hep-ph]}}.

\bibitem{Choi:1996fs}
K.~Choi, H.~B. Kim, and J.~E. Kim, ``{Axion cosmology with a stronger QCD in
  the early universe},''
  \href{http://dx.doi.org/10.1016/S0550-3213(97)00066-7}{{\em Nucl. Phys. B}
  {\bfseries 490} (1997) 349--364},
  \href{http://arxiv.org/abs/hep-ph/9606372}{{\ttfamily arXiv:hep-ph/9606372}}.

\bibitem{Co:2018phi}
R.~T. Co, E.~Gonzalez, and K.~Harigaya, ``{Axion Misalignment Driven to the
  Bottom},'' \href{http://dx.doi.org/10.1007/JHEP05(2019)162}{{\em JHEP}
  {\bfseries 05} (2019) 162}, \href{http://arxiv.org/abs/1812.11186}{{\ttfamily
  arXiv:1812.11186 [hep-ph]}}.

\bibitem{Matsui:2020wfx}
H.~Matsui, F.~Takahashi, and W.~Yin, ``{QCD Axion Window and False Vacuum Higgs
  Inflation},'' \href{http://dx.doi.org/10.1007/JHEP05(2020)154}{{\em JHEP}
  {\bfseries 05} (2020) 154}, \href{http://arxiv.org/abs/2001.04464}{{\ttfamily
  arXiv:2001.04464 [hep-ph]}}.

\bibitem{Guth:2018tdu}
F.~Takahashi, W.~Yin, and A.~H. Guth, ``{QCD axion window and low-scale
  inflation},'' \href{http://dx.doi.org/10.1103/PhysRevD.98.015042}{{\em Phys.
  Rev. D} {\bfseries 98} no.~1, (2018) 015042},
  \href{http://arxiv.org/abs/1805.08763}{{\ttfamily arXiv:1805.08763
  [hep-ph]}}.

\bibitem{DiLuzioAxionLandscape}
L.~Di~Luzio, M.~Giannotti, E.~Nardi, and L.~Visinelli, ``{The landscape of QCD
  axion models},'' \href{http://dx.doi.org/10.1016/j.physrep.2020.06.002}{{\em
  Phys. Rept.} {\bfseries 870} (2020) 1--117},
  \href{http://arxiv.org/abs/2003.01100}{{\ttfamily arXiv:2003.01100
  [hep-ph]}}.

\bibitem{AAdI}
G.~Dvali and J.~B. Gómez, ``Axion alignment during inflation,'' unpublished
  Thesis by J. B. Gómez.

\bibitem{PQMechanism}
R.~D. Peccei and H.~R. Quinn, ``Constraints imposed by $\mathrm{CP}$
  conservation in the presence of pseudoparticles,''
  \href{http://dx.doi.org/10.1103/PhysRevD.16.1791}{{\em Phys. Rev. D}
  {\bfseries 16} (Sep, 1977) 1791--1797}.
  \url{https://link.aps.org/doi/10.1103/PhysRevD.16.1791}.

\bibitem{WeinbergAxion}
S.~Weinberg, ``A new light boson?''
  \href{http://dx.doi.org/10.1103/PhysRevLett.40.223}{{\em Phys. Rev. Lett.}
  {\bfseries 40} (Jan, 1978) 223--226}.
  \url{https://link.aps.org/doi/10.1103/PhysRevLett.40.223}.

\bibitem{WilczekAxion}
F.~Wilczek, ``{Problem of Strong $P$ and $T$ Invariance in the Presence of
  Instantons},''
\href{http://dx.doi.org/10.1103/PhysRevLett.40.279}{{\em Phys. Rev. Lett.}
  {\bfseries 40} (1978) 279--282}.

\bibitem{PhysRevD.17.2717}
C.~G. Callan, R.~Dashen, and D.~J. Gross, ``Toward a theory of the strong
  interactions,'' \href{http://dx.doi.org/10.1103/PhysRevD.17.2717}{{\em Phys.
  Rev. D} {\bfseries 17} (May, 1978) 2717--2763}.
  \url{https://link.aps.org/doi/10.1103/PhysRevD.17.2717}.

\bibitem{QCDINstantonsFiniteTemp}
D.~J. Gross, R.~D. Pisarski, and L.~G. Yaffe, ``Qcd and instantons at finite
  temperature,'' \href{http://dx.doi.org/10.1103/RevModPhys.53.43}{{\em Rev.
  Mod. Phys.} {\bfseries 53} (Jan, 1981) 43--80}.
  \url{https://link.aps.org/doi/10.1103/RevModPhys.53.43}.

\bibitem{Hertzberg:2008wr}
M.~P. Hertzberg, M.~Tegmark, and F.~Wilczek, ``{Axion Cosmology and the Energy
  Scale of Inflation},''
  \href{http://dx.doi.org/10.1103/PhysRevD.78.083507}{{\em Phys. Rev. D}
  {\bfseries 78} (2008) 083507},
  \href{http://arxiv.org/abs/0807.1726}{{\ttfamily arXiv:0807.1726
  [astro-ph]}}.

\bibitem{Kolb:2003ke}
E.~W. Kolb, A.~Notari, and A.~Riotto, ``{On the reheating stage after
  inflation},'' \href{http://dx.doi.org/10.1103/PhysRevD.68.123505}{{\em Phys.
  Rev. D} {\bfseries 68} (2003) 123505},
  \href{http://arxiv.org/abs/hep-ph/0307241}{{\ttfamily arXiv:hep-ph/0307241}}.

\bibitem{Planck2018}
{\bfseries Planck} Collaboration, N.~Aghanim {\em et~al.}, ``{Planck 2018
  results. VI. Cosmological parameters},''
\href{http://arxiv.org/abs/1807.06209}{{\ttfamily arXiv:1807.06209
  [astro-ph.CO]}}.

\bibitem{Marsh:2015xka}
D.~J.~E. Marsh, ``{Axion Cosmology},''
  \href{http://dx.doi.org/10.1016/j.physrep.2016.06.005}{{\em Phys. Rept.}
  {\bfseries 643} (2016) 1--79},
  \href{http://arxiv.org/abs/1510.07633}{{\ttfamily arXiv:1510.07633
  [astro-ph.CO]}}.

\bibitem{Kamada:2014ufa}
K.~Kamada, ``{Inflationary cosmology and the standard model Higgs with a small
  Hubble induced mass},''
  \href{http://dx.doi.org/10.1016/j.physletb.2015.01.024}{{\em Phys. Lett. B}
  {\bfseries 742} (2015) 126--135},
  \href{http://arxiv.org/abs/1409.5078}{{\ttfamily arXiv:1409.5078 [hep-ph]}}.

\bibitem{Dvali:2021byy}
G.~Dvali, F.~K\"uhnel, and M.~Zantedeschi, ``{Primordial black holes from
  confinement},'' \href{http://dx.doi.org/10.1103/PhysRevD.104.123507}{{\em
  Phys. Rev. D} {\bfseries 104} no.~12, (2021) 123507},
  \href{http://arxiv.org/abs/2108.09471}{{\ttfamily arXiv:2108.09471
  [hep-ph]}}.

\bibitem{PDG2020}
{\bfseries Particle Data Group} Collaboration, P.~Zyla {\em et~al.}, ``{Review
  of Particle Physics},'' \href{http://dx.doi.org/10.1093/ptep/ptaa104}{{\em
  PTEP} {\bfseries 2020} no.~8, (2020) 083C01}.

\bibitem{Bjorkeroth:2019jtx}
F.~Bj\"orkeroth, L.~Di~Luzio, F.~Mescia, E.~Nardi, P.~Panci, and R.~Ziegler,
  ``{Axion-electron decoupling in nucleophobic axion models},''
  \href{http://dx.doi.org/10.1103/PhysRevD.101.035027}{{\em Phys. Rev. D}
  {\bfseries 101} no.~3, (2020) 035027},
  \href{http://arxiv.org/abs/1907.06575}{{\ttfamily arXiv:1907.06575
  [hep-ph]}}.

\bibitem{ELLIS1979141}
J.~Ellis and M.~K. Gaillard, ``Strong and weak cp violation,''
  \href{http://dx.doi.org/https://doi.org/10.1016/0550-3213(79)90297-9}{{\em
  Nuclear Physics B} {\bfseries 150} (1979) 141--162}.
  \url{https://www.sciencedirect.com/science/article/pii/0550321379902979}.

\bibitem{Jeong:2013xta}
K.~S. Jeong and F.~Takahashi, ``{Suppressing Isocurvature Perturbations of QCD
  Axion Dark Matter},''
  \href{http://dx.doi.org/10.1016/j.physletb.2013.10.061}{{\em Phys. Lett. B}
  {\bfseries 727} (2013) 448--451},
  \href{http://arxiv.org/abs/1304.8131}{{\ttfamily arXiv:1304.8131 [hep-ph]}}.

\end{thebibliography}\endgroup


\providecommand{\href}[2]{#2}\begingroup\raggedright\endgroup

\end{document}